\documentclass[a4paper]{article}

\usepackage[latin1]{inputenc}
\usepackage{amsmath}

\usepackage{wrapfig}

\usepackage[english]{babel}
\usepackage{amssymb}
\usepackage{indentfirst}

\usepackage{ae}

\usepackage[pdftex]{graphicx}

\usepackage[pdftex]{geometry}
\title{\textsf{Relativistically speaking: Let's walk or run through the rain?}}

\author{\\
\vspace{-0.9cm}
\\
Armando V.D.B. Assis\thanks{Dedicado a YHVH.}\\
  \texttt{armando.assis@pgfsc.ufsc.br}
}
\date{\today}

\begin{document}

\maketitle

\vspace{-0.7cm}
\begin{abstract}
  We analyse under a simple approach the problem one must decide the best strategy to minimize the contact with rain when moving between two points through the rain. The avaiable strategies: walk (low speed boost $<<$ $c$) or run (relativistic speed boost $\approx$ $c$).
\end{abstract}
\section{On the General Aspects of the Problem}

We will restrict ourselves to the domain of relativistic kinematics. Consider an inertial cartesian reference frame $0xyz$. Put the velocity field $\vec{u}\left(\vec{r};t\right)\equiv\vec{u}\left(x,y,z;t\right)$ of the raindrops observed in the inertial reference frame $0xyz$. Let a closed surface $\mathcal{S}$ move\footnote{Each point belonging to $\mathcal{S}$ is moving with the same constant velocity $\vec{v}$ in $0xyz$.} with constant velocity $\vec{v}$ in $0xyz$. Now, let $0'x'y'z'$ be the inertial reference frame encrusted to the proper surface $\mathcal{S}^{'}$ and $\vec{u}'\left(\vec{r}';t'\right)\equiv\vec{u}'\left(x',y',z';t'\right)$ the velocity field of the raindrops observed in $0'x'y'z'$. Suppose $\mathcal{S}^{'}$ as being transparent to the field $\vec{u}'\left(\vec{r}';t'\right)$, i.e., an imaginary proper control surface. Therefore, the flux $\phi'$ of the velocity field $\vec{u}'\left(x',y',z';t'\right)$ through $\mathcal{S}^{'}$ is given by:\\
\begin{equation}
\phi'=\oint_{\mathcal{S}^{'}}\vec{u}'\left(x',y',z';t'\right)\cdot\hat{n}dS',
\end{equation}\\
where $\hat{n}dS'$ is the elementar surface vector normal to $\mathcal{S}^{'}$ at a point $P\left(x',y',z';t'\right)$ $\in$ $\mathcal{S}^{'}\times\left\{t'\right\}$. Obviously, a real body moving through the rain will be opaque to the field $\vec{u}'\left(\vec{r}';t'\right)$. In cases in which there exists a shadow region regarding the surfaces where the field $\vec{u}'\left(\vec{r}';t'\right)$ is null, the amount of rain that would penetrate through the surfaces of the opaque body in a case of transparency is given by eq. (1):\\
\begin{displaymath}
\phi'=\oint_{\mathcal{S^{'}}}\vec{u}'\left(x',y',z';t'\right)\cdot\hat{n}dS'=\int_{\mathcal{S}^{'}_{r}}\vec{u}'_{r}\left(x',y',z';t'\right)\cdot\hat{n}dS'+\int_{\mathcal{S}^{'}_{\textit{null}}}\vec{0}\cdot\hat{n}dS'\therefore
\end{displaymath}
\begin{equation}
\phi'=\phi'_{r}=\int_{\mathcal{S}_{r}}\vec{u}'_{r}\left(x',y',z';t'\right)\cdot\hat{n}dS',
\end{equation}\\
where $\phi'_{r}$ is the flux that would penetrate (in a case of transparency) through the region of $\mathcal{S}^{'}$ touching raindrops, namely  $\mathcal{S}^{'}_{r}$. Hence, eq. (2) gives the amount of water touching $\mathcal{S}^{'}$ through $\mathcal{S}^{'}_{r}\subseteq\mathcal{S}^{'}$. $\mathcal{S}^{'}_{\textit{null}}$ is the part of $\mathcal{S^{'}}$ at shadow region where $\vec{u}'=\vec{0} \; \; \forall$ $P\left(x',y',z';t'\right)$ $\in$ $\mathcal{S}^{'}_{\textit{null}}\times\{t'\}$.
\section{A Simple Approach}
\vspace{-1.5cm}
\begin{wrapfigure}[11]{l}{100pt}
\centering
\includegraphics[scale=.27]{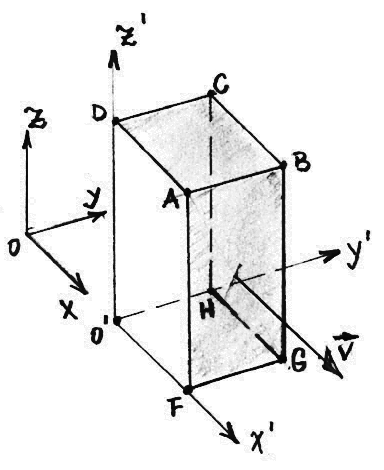}
\end{wrapfigure}
\vspace{1.7cm}
Let's do some simplifications in analyzing the problem. The surface $\mathcal{S}'$ will be of a proper (there is not Lorentz contraction regarding $\mathcal{S}^{'}$) rectangular prism with square base in $0'x'y'z'$, as drawn on the left. This schematic drawing shows the mentioned reference frames, $0xyz$ and $0'x'y'z'$, and the boost vector $\vec{v}$. We also note the surface $DAFGBCD$, taken as the surface $\mathcal{S}^{'}_{r}$ touching the raindrops (these ones are not depicted yet).

In the schematic drawing below, we have a lateral vision of the problem as observed in $0xyz$, with the terminal velocity field of the raindrops $\vec{u}(x,y,z;t)$, the position vectors $\vec{r}$ and $\vec{r}'$, of an arbitrary rain drop $P$, the displacement vector of the origin $0'$ in relation to the origin $0$, vector $\vec{h}$ and other trivially denoted elements inherent to the simplification we are from now on considering.\newpage

Let's consider the two reference frames in the canonical configuration, i.e., coincident origins at $t=t'=0$ keeping the spacelike parallelism of the axes $x\equiv x'$, $y\equiv y'$ and $z\equiv z'$.\\
\vspace{-0.3cm}
\begin{wrapfigure}[19]{l}{185pt}
\centering
\includegraphics[scale=.25]{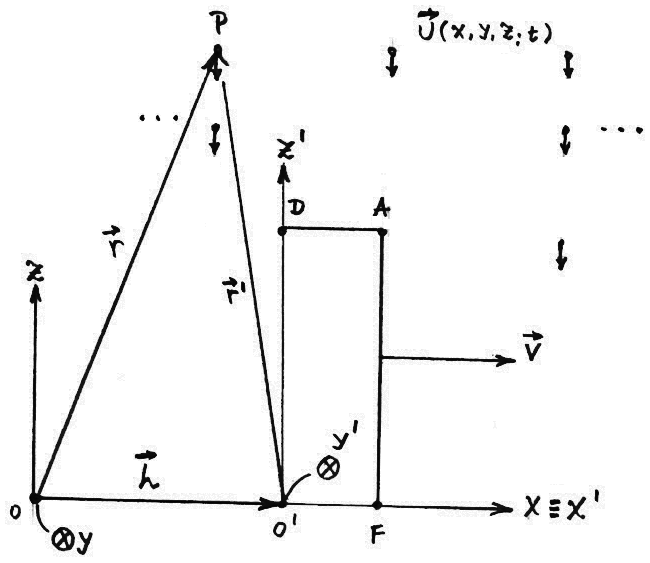}
\end{wrapfigure}
The Lorentz transformations of the 3-velocity field of the raindrops gives $\vec{u}'=d\vec{r}'/dt'$ in the $0x'y'z'$ reference frame. Since $(u_{x},u_{y},u_{z})=(0,0,-u_{\infty})$ in $0xyz$, where $u_{\infty}$ is the terminal velocity of the raindrops as measured in $0xyz$, we have:\\
\vspace{-0.7cm}
\\  
\begin{displaymath}
u_{x'}=\frac{u_{x}-v}{1-vu_{x}/c^{2}}=-v;\,\,\,\,\,\,\,\,\,\,\,\,\,\,\,\,
\end{displaymath}
\begin{displaymath}
u_{y'}=\frac{u_{y}}{\gamma\left(1-vu_{x}/c^{2}\right)}=0;\,\,\,\,\,\,\,\,\,\,\,\,
\end{displaymath}
\begin{displaymath}
u_{z'}=\frac{u_{z}}{\gamma\left(1-vu_{x}/c^{2}\right)}=-\frac{u_{\infty}}{\gamma},
\end{displaymath}\\
\vspace{-0.2cm}
\\
where $\gamma=1/\sqrt{1-v^{2}/c^{2}}$, $c$ is the speed of light. We tacitly started supposing the uniform field of the terminal raindrops as a simplification. As mentioned above, we are concerned to the kinematics, in which case such a constraint is sufficient to the flux calculation, regardless the physics of drag opon the raindrops and so on. We must discretize the field, since we should not associate a vector to every point of space $0xyz$ in virtue of the fact we are not dealing with continuous current lines of fluid.
Putting the above results in eq. (2), asseverating discreteness by the label \textit{d}, we have:\\
\vspace{-0.4cm}
\\
\begin{equation}
\phi'_{r}=\int_{DAFGBCD}\left(-v\hat{e}_{x'}-\frac{u_{\infty}}{\gamma}\hat{e}_{z'}\right)_{\textit{d}}\cdot\hat{n}dS',
\end{equation}\\
\vspace{-0.2cm}
\\
where $\hat{e}_{x'}$ and $\hat{e}_{z'}$ are the unitary vectors along the axes $0'x'$ and $0'z'$ respectively.
 
We discretize the raindrops field at the points of the proper surfaces DABCD and AFGBA in the $0'x'y'z'$ reference frame via the Dirac delta functions at the points of these surfaces that simultaneously cross the discrete rain rays. Obviously, these non-proper surface points in $0xyz$ cross the discrete rain rays non-simultaneously in $0xyz$. It does not matter at all, since we are analyzing the problem under $0'x'y'z'$ point of view. In fact, the rain rays appears to be rotated when compared to the non-relativistic case in which the angle $\theta$ between the rain drops ray and the $0'x'$ axe is given by $\tan{\theta}=u_{\infty}/v$, instead of $\tan{\theta}=u_{\infty}/(\gamma v)$. Hence, simultaneous raindrops under non-relativistic case are shifted through the surfaces, becoming non-simultaneous ones. The non-null raindrops field at the surfaces DABCD and AFGBA at an instant $t'$ in $0'x'y'z'$ world description reads:\\
\begin{equation}
\left(-v\hat{e}_{x'}-\frac{u_{\infty}}{\gamma}\hat{e}_{z'}\right)_{\textit{ABCDA}}=\left(-v\hat{e}_{x'}-\frac{u_{\infty}}{\gamma}\hat{e}_{z'}\right)\displaystyle\sum_{i=1}^{N_{x'}(t')}\sum_{j=1}^{N_{y'}(t')}\int_{-\infty}^{\infty}\int_{-\infty}^{\infty}\delta\left[x'-x_{i}'(t')\right]\delta\left[y'-y_{j}'(t')\right] \; dx'dy';
\end{equation}
\begin{equation}
\left(-v\hat{e}_{x'}-\frac{u_{\infty}}{\gamma}\hat{e}_{z'}\right)_{\textit{ABGFA}}=\left(-v\hat{e}_{x'}-\frac{u_{\infty}}{\gamma}\hat{e}_{z'}\right)\displaystyle\sum_{j=1}^{N_{y'}(t')}\sum_{k=1}^{N_{z'}(t')}\int_{-\infty}^{\infty}\int_{-\infty}^{\infty}\delta\left[y'-y_{i}'(t')\right]\delta\left[z'-z_{j}'(t')\right] \; dy'dz',
\end{equation}\\
with $N_{x'_{j}}(t')\geq 1$ $\forall j$ $\in$ $\{1,\,2,\,3\}$, where $N_{x'}(t')N_{y'}(t')=N_{\textit{ABCDA}}(t')$ is the number of raindrops simultaneously touching the proper surface ABCDA at the instant $t'$ and $N_{y'}(t')N_{z'}(t')=N_{\textit{ABGFA}}(t')$ is the number of raindrops simultaneously touching the proper surface ABGFA at the instant $t'$, both measured in $0'x'y'z'$.

By the schematic drawing below, we depict the situation under $0'x'y'z'$ point of wiew. We denote $u\equiv u_{\infty}/\gamma$. From the depicted geometry, we infer the following relations:
\begin{equation}
SA=AS\cong\frac{DA}{N_{x'}},\,\,\,\,\,AT\cong\frac{AF}{N_{z'}},\,\,\,\,\,\frac{AT}{AS}=\frac{AF}{DA}\frac{N_{x'}}{N_{z'}}\,\,\,\,\,\therefore
\end{equation}
\\
\vspace{-0.7cm}
\\
\begin{equation}
\frac{AT}{AS}=\frac{AF}{DA}\frac{N_{x'}}{N_{z'}}\frac{N_{y'}AB}{N_{y'}AB}.
\end{equation}
$N_{x'}N_{y'}=N_{\textit{ABCDA}}$ is the number of raindrops touching the proper surface ABCDA, from now denoted\footnote{$s$ denotes \textit{superior} and $f$ denotes \textit{frontal}.} by $N_{s}$, and $N_{y'}N_{z'}=N_{\textit{ABGFA}}$ the number of raindrops touching the proper surface ABGFA, from now denoted by $N_{f}$. 
\newpage
\begin{wrapfigure}[26]{l}{240pt}
\centering
\includegraphics[scale=.62]{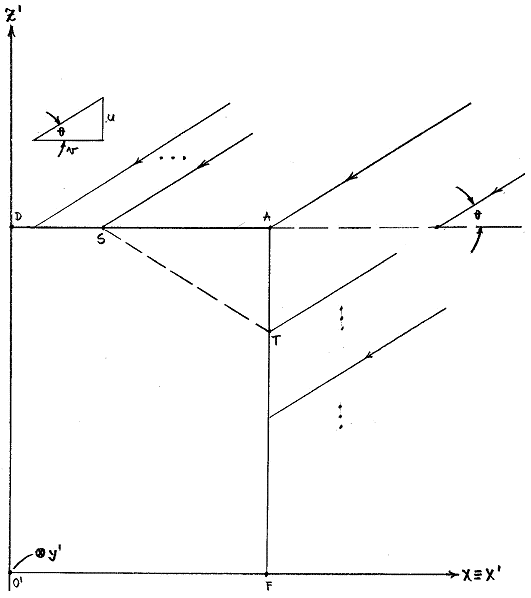}
\end{wrapfigure}
Since $AF\cdot AB$ is the proper area of the frontal surface ABGFA, namely $A_{f}$, $DA\cdot AB$ the proper area of the top (superior) surface ABCDA, namely $A_{s}$, eq. (7) reads:\\
\begin{displaymath}
\frac{AT}{AS}=\frac{A_{f}}{A_{s}}\frac{N_{s}}{N_{f}}\therefore
\end{displaymath}
\begin{equation}
N_{s}=\frac{AT}{AS}\frac{A_{s}}{A_{f}}N_{f}.
\end{equation}\\
Simultaneous raindrops touching the top proper surface in $0'x'y'z'$ and simultaneous raindrops touching the frontal proper surface in $0'x'y'z'$ are not necessarily simultaneously simultaneous in both surfaces, altough eqs. (6), (7) and (8) remain  \textit{geometrically } holding anyway: one would only have to shift forward/backward the frontal proper surface to recover the above \textit{geometrically} depicted situation.. Hence, in virtue of discreteness, one writes down the instantaneous flux:\\
\vspace{0.1cm}
\\
\begin{equation}
N_{x'}(t')N_{y'}(t')\,a_{ij}^{s}(t')\frac{u_{\infty}}{\gamma}=\sum_{k=0}^{I(\tau'/T_{z'})} V_{ij}^{s}(t')\,N_{x'}N_{y'}\delta\left(t'-kT_{z'}\right);\,\,\,\,\,\,\,\,\,\,\,\,\,\,\,\,\,
\end{equation}\\
\vspace{-0.5cm}
\\
\begin{equation}
N_{y'}(t')N_{z'}(t')\,a_{jm}^{f}(t')v=\sum_{l=0}^{I(\tau'/T_{x'})} V_{jm}^{f}(t')\,N_{y'}N_{z'}\delta\left(t'-lT_{x'}\right);\,\,\,\,\,\,\,\,\,\,\,\,
\end{equation}\\
\vspace{-0.4cm}
\\
\begin{equation}
N_{x'}(t')N_{y'}(t')=N_{s}(t');\,\,\,N_{y'}(t')N_{z'}(t')=N_{f}(t'),
\end{equation}
\\
\vspace{-0.3cm}
\\
where the function $I(\tau'/T_{x'_{j}}):\tau'/T_{x'_{j}}\in\mathbb{R}^{+}\rightarrow I(\tau'/T_{x'_{j}})\in\mathbb{Z}^{+}$ returns the integer part of $\tau/T_{x'_{j}}$, the ratio between the total elapsed time $\tau'$ in $0'x'y'z'$ world and the period $T_{x'_{j}}$ of a plane orthogonal to $0'x'_{j}$ containing simultaneous raindrops in $0'x'y'z'$ world; $\delta$ is the Dirac delta function; $N_{x'_{j}}$ was properly defined through the march that led to eqs. (6), (7) and (8) above. Furthermore, since $\delta$ is a distribution, the inherent context regarding it will become clear in the time integration we will perform at the final of our calculation. $N_{s}(t')$ and $N_{f}(t')$ are, respectively, the $t'$ instantaneous number of raindrops touching the top and the frontal proper surfaces of the prism in $0'x'y'z'$ world. $a_{ij}^{s}(t')$ is the $0'z'$ orthogonally projected area of a single raindrop $t'$ instantaneously passing through the superior (top) proper adjacent surface of the prism at some $(i,j)$-position, $V_{ij}^{s}(t')$ is the $t'$ instantaneous raindrop volume passing through
$a_{ij}^{s}(t')$, both observed in $0'x'y'z'$ world; $a_{jm}^{f}(t')$ and $V_{jm}^{f}(t')$ have analogous definitions related to the frontal proper surface of the prism.

We asseverate that the identity of raindrops touching the frontal proper surface ABGFA, $N_{f}$, and the identity of raindrops touching de top surface ABCDA, $N_{s}$, must be the same in both reference frames, but not simultaneous in both systems of reference. These observers will not agree about the identities of the raindrops simultaneously touching the top and frontal surfaces. This lack of reciprocal simultaneity, an inherent character of Einstein's theory of relativity, tell us we must avoid a measure of the number of raindrops based on identities of raindrops simultaneously touching the surfaces in both frames of reference, as naturally one would do in a case of galilean relativity. Straightforwardly, imagine (at $0xyz$!) a set of horizontal planes equally spaced ($d_{0z}$) (each containing horizontal lattice of raindrops $0x\times 0y$ spaced) falling vertically at constant terminal velocity $u$. The prism touches a set of vertical planes periodically with period $d_{0x}/v$. At each instant, the vertical displacement between consecutive raindrops touching the frontal surface of the prism is a constant, say $d_{0z}$. The distance $AT$ is $d_{0z}$ as easily verified in the above schematic instantaneous. Indeed, the vertical raindrops touching the frontal surface do that simultaneously in each frame of reference, since two subsequent raindrops in a vertical plane parallel to $yz$ simultaneously touch the frontal surface as observed in $0xyz$, say at $t_{1}=t_{2}$, with $x_{1}=x_{2}$ (same vertical plane $yz$ in $0xyz$), and one easily verify by Lorentz transformation of the instants of these events in $0'x'y'z'$:
\begin{equation}
t'_{2}-t'_{1}=\gamma\left(t_{2}-\frac{vx_{2}}{c^{2}}\right)-\gamma\left(t_{1}-\frac{vx_{1}}{c^{2}}\right)=0;
\end{equation}
since $z'_{2}-z'_{1}=z_{2}-z_{1}$ for any pair of events under canonical Lorentz transformation, $AT=d_{0z}$. Unfortunatelly, the simultaneity of raindrops touching the top area in $0xyz$ does not hold in $0'x'y'z'$, since $x_{2}\neq x_{1}$ for any pair of raindrops contained in the $xy$ plane simultaneously touching the top area as observed in $0xyz$, although $t_{2}=t_{1}$. Hence, since we are considering simultaneity in $0'x'y'z'$ in the situation depicted in the above schematic drawing, the proper top plane events in $0'x'y'z'$ like A, S etc., at an instant $t'$ must be under the spacetime constraint condition in $0xyz$ (spacetime set of points correlated to $t'$):
\begin{equation}
t'=\textit{constant}=\gamma\left(t-\frac{vx}{c^{2}}\right),
\end{equation}\\
\vspace{-0.5cm}
\\
where a particular value of the constant $t'$ defines a particular set of spacetime points coordinated in $0xyz$ simultaneous in $0'x'y'z'$ at $t'$ as measured by sincronous $0'x'y'z'$ clocks. These spacetime points in $0xyz\times t$ must have diferent heights in $0z$. Two of such points are $z$ shifted at $t_{1}<t_{2}$:
\begin{equation}
\gamma\left(t_{2}-\frac{vx_{2}}{c^{2}}\right)=\gamma\left(t_{1}-\frac{vx_{1}}{c^{2}}\right)\Rightarrow t_{2}-t_{1}=\frac{v}{c^{2}}\left(x_{2}-x_{1}\right)\,\,\,\therefore\,\,\,\frac{v}{c^{2}}n_{x}d_{0x}=\frac{n_{z}d_{0z}}{u_{\infty}},
\end{equation}
$n_{x}$, $n_{z}$ integers characterizing the $0x$ and $0z$ shifts of these pair of arbitrary points belonging to some of the hypersurface families given by eq. (13). At the instant $t'$ depicted in the above drawing, a pair of points like A, S etc., simultaneously touching the top surface in $0'x'y'z'$ reference frame, are $x'$ simultaneous spaced by an amount given by the Lorentz tranformation of the correlated two points in $0xyz$ spaced (in $0xyz$ reference frame) by an amount given by eq. (14) Hence:
\begin{equation}
x'_{2}-x'_{1}=\gamma\left(x_{2}-vt_{2}\right)-\gamma\left(x_{1}-vt_{1}\right)=\gamma\left[\left(x_{2}-x_{1}\right)-v\left(t_{2}-t_{1}\right)\right]\stackrel{(14)}{=}\gamma\left(1-\frac{v^{2}}{c^{2}}\right)n_{x}d_{0x}=\gamma^{-1}n_{x}d_{0x}.
\end{equation}
Hence, consecutive points simultaneously touching the proper top surface of the prism in $0'x'y'z'$ reference frame are spaced apart by the amount given by $n_{x}=1$ in eq. (15)\footnote{See the meaning of $AS$ in eqs. (6), (7) and (8).}:
\begin{equation}
AS=\gamma^{-1}d_{0x},
\end{equation}
as one would intuitivelly expect as being the Lorentz contraction of the proper displacement $d_{0x}$. Unfortunatelly, one should take some care in defining simultaneous horizontal raindrops in $0'x'y'z'$ when seeking the instantaneous flux calcutation in relativistic (Einstein's) discrete situations. Strictly speaking, in virtue of eq. (14), such simultaneity occurs in $0'x'y'z'$ only if:
\begin{equation}
n_{z}=\frac{u_{\infty}v}{c^{2}}\frac{d_{0x}}{d_{0z}},
\end{equation}
for consecutive points, with $n_{z}$ $\in$ $\mathbb{Z}$. In a galilean case, $uv<<c^{2}\Rightarrow n_{z}d_{0z}\approx 0$, and there would not exist the necessity of a $z$ shift condition to guarantee a top proper plane of simultaneity in $0'x'y'z'$. Physically, the vertical velocity of the raindrops $u_{\infty}$ is the cause of the absence of a general simetry regarding the usage of simultaneity for purposes of instantaneous flux calculation. In a case in wich one was able to adhocally consider continuity, one would straighforwardly use simultaneous reasoning in $0'x'y'z'$ ab ibitio, since such would imply:
\begin{equation}
n_{x}d_{0x}\rightarrow dx,\,\,\,\,\,n_{z}d_{0z}\rightarrow dz,
\end{equation}\\
\vspace{-0.5cm}
\\
but this is not, until now, the case being analysed here. One shall infer that simultaneous points in $0'x'y'z'$ belonging to the respective $x'y'$ plane parallel to the top surface of the prism must move diagonally in $0'x'y'z'$, must move downwards with velocity $u_{\infty}/\gamma$ and backwards with velocity $-v$, since these are the velocity components of these points in $0'x'y'z'$. Considering one of such horizontal plane of simultaneous (if possible) raindrops in $0'x'y'z'$, one also shall infer from eq. (14) that the respective raindrops in $0xyz$ must be diagonally located through sucessive horizontal $xy$ planes of simultaneity in $0xyz$, being $\alpha=\arctan{u_{\infty}v/c^{2}}$ the angle between this diagonal and $0x$. These diagonals move downwards with velocity $u_{\infty}$ while the prism moves forward with velocity $v$ in $0xyz$ world. Each diagonal corresponds to a $x'y'$ plane of raindrops simultaneously touching the proper top plane of the prism in $0'x'y'z'$ world. The interpretation is simple: since raindrops touch the top surface of the prism simultaneously in $0xyz$ world, they do not simultaneously touch in $0'x'y'z'$ world. This lack of simultaneity is asseverated by the necessity $n_{z}<<1$ as stated by eq. (17).

From now on, we model our raindrops distribution from the perspective of an $0xyz$ observer, as one would naturally do, with the following characteristcs:
\begin{itemize}
\item The average proper displacement of raindrops along $0x$ is $d_{0x}$;
\item The average proper displacement of raindrops along $0y$ is $d_{0y}$;
\item The average proper displacement of raindrops along $0y$ is $d_{0y}$;
\item We will neglect relativistic (Einstein's) effects of raindrops in $0xyz$;
\item The raindrops' average bahavior in $0xyz$ will be translated to a tridimensional raindrops $d_{0x}\times d_{0y}\times d_{0z}$ othogonally spaced infinite lattice falling at terminal velocity $u_{\infty}$, being the sites' basis vectors given by $\{\vec{d}_{0x}=d_{0x}\hat{e}_{x},\,\,\vec{d}_{0y}=d_{0y}\hat{e}_{y},\,\,\vec{d}_{0z}=d_{0z}\hat{e}_{z}\}$, where $\{\hat{e}_{x},\,\,\hat{e}_{y},\,\,\hat{e}_{z}\}$ is the canonical spacelike 3D euclidian orthonormal basis of $0xyz$.
\end{itemize} 
\section{Solving the Problem}

Hence, the contact with the top plane of the prism is simultaneous in the $0xyz$ world, implying non-simultaneity of these raindrops in the $0'x'y'z'$ world. The distribution of these raindrops must have, \textit{instantaneously} at $t'$ in $0'x'y'z'$ world, the following characteristics:
\begin{itemize}
\item The displacement between two consecutive raindrops correlated to the respective simultaneous ones in $0xyz$, these latter displaced by the proper distance $x_{i+1}-x_{i}=d_{0x}$ along $0x$ and belonging to the falling $xy$ plane in $0xyz$, is given by:
\begin{equation}
x'_{i+1}(t')-x'_{i}(t')=\gamma^{-1}\left(x_{1+1}-x_{i}\right)=\gamma^{-1}d_{0x}.
\end{equation}
\item The displacement between two consecutive raindrops correlated to the respective simultaneous ones in $0xyz$, these latter displaced by the proper distance $z_{i+1}-z_{i}=0$ along $0z$ and belonging to the falling $xy$ plane in $0xyz$, is given by:\\
\vspace{-0.7cm}
\\
\begin{equation}
z'_{i}(t')-z'_{i+1}(t')=\frac{u_{\infty} v d_{0x}}{c^{2}}.
\end{equation}
\item The distance between consecutive raindrop planes $\Pi_{i+1}$ and $\Pi_{i}$, $\forall$ $i$, is $d_{0z}$. The raindrop planes are inclined in relation to the $x'y'$ plane by the angle:
\begin{equation}
\alpha=\pi-\arctan{\left(\frac{\gamma u_{\infty}v}{c^{2}}\right)}.
\end{equation} 
\end{itemize}
Indeed, let's derive these facts. Firstly, instantaneously at $t$ in $0xyz$, two consecutive raindrops $0x$ along, are time delayed in $0'x'y'z'\times\{t'\}$ world by the amount:
\begin{equation}
t'_{i+1}-t'_{i}=\gamma\left(t-\frac{v}{c^{2}}x_{i+1}\right)-\gamma\left(t-\frac{v}{c^{2}}x_{i}\right)=-\gamma\frac{v}{c^{2}}\left(x_{i+1}-x_{i}\right)=-\gamma\frac{v}{c^{2}}d_{0x},
\end{equation}
and the $i$-raindrop is late in relation to the $(i+1)$-raindrop. Hance, backwarding the $t'_{i}$ clocks down to the the $t'_{i+1}$ instant, the $i$-raindrop must move the amounts: $\delta z'$ upwards and $\delta x'$ to the right, being these amounts given by:
\begin{equation}
\delta z'=\left(-\frac{u_{\infty}}{\gamma}\right)\times\left(-\gamma\frac{v}{c^{2}}d_{0x}\right)=\frac{u_{\infty} v d_{0x}}{c^{2}};\,\,\,\delta x'=\left(-v\right)\times\left(-\gamma\frac{v}{c^{2}}d_{0x}\right)=\frac{v^{2}\gamma d_{0x}}{c^{2}},
\end{equation}
since $\left(-v\hat{e}_{x'}-(u_{\infty}/\gamma)\hat{e}_{z'}\right)$ is the velocity of raindrops in $0'x'y'z'$. But, at $t$, the $i$-raindrop and the $(i+1)$-raindrop have got the same $z$ coordinate, since they are in a $xy$ plane, and, since the $z\rightarrow z'$ Lorentz map is identity, these raindrops must have the same $z'$ coordinate at their respective transformed instants. Hence, backwarding $t'_{i}$ clocks down to the the $t'_{i+1}$ instant, one concludes that the $\delta z'$ in eq. (21) is the instantaneous, at same $t'$, height shift between consecutive raindrops that simultaneously touches the top plane of the prism in $0xyz$. The $x\rightarrow x'$ Lorentz map is not identity, implying one must calculate the $x'_{i+1}-x'_{i}$ shift at the $0xyz$ instantaneous $t$:
\begin{equation}  
x'_{i+1}(t)-x'_{i}(t)=\gamma\left(x_{i+1}-vt\right)-\gamma\left(x_{i}-vt\right)=\gamma\left(x_{i+1}-x_{i}\right)=\gamma d_{0x}.
\end{equation}
Hence, backwarding $t'_{i}$ clocks down to the the $t'_{i+1}$ instant, this amount is reduced by the amount $\delta x'$ given by eq. (22):
\begin{equation}
x'_{i+1}(t')-x'_{i}(t')=\gamma d_{0x}-\gamma d_{0x}\frac{v^{2}}{c^{2}}=\gamma d_{0x}\left(1-\frac{v^{2}}{c^{2}}\right)=\gamma^{-1}d_{0x}.
\end{equation}
From eqs. (23) and (24) we reach the eqs. (19) and (20). One shall infer that the non-instantaneous displacement (non-instantaneous in $0'x'y'z'$) given by eq. (24) is the distance between two sucessive non-instantaneous raindrops marks assigned upon the proper top plane of the prism in $0'x'y'z'$. This fact is easy to understand, as these instantaneously assigned marks (instantaneous in $0xyz$) become splayed in $0'x'y'z'$, since the prism turns out to be Lorentz contracted in $0xyz$. Also, one shall infer that eq. (25) gives the $t'$ instantaneous displacement of falling raindrops along $0'x'$. The reason why the distance between consecutive raindrops marks $\gamma d_{0x}$ are bigger than the contracted distance $\gamma^{-1}d_{0x}$ of the two consecutive falling raindrops is explained by the non-simultaneity between these raindrops when touching the proper top plane of the prism in the $0'x'y'z'$ world, straightforwardly seem by the inclination between the raindrop plane containing these two consecutives raindrops and the proper top plane of the prism; i.e., when the first raindrop touches the prism at the proper top plane assigning the first mark, the second travels an amount $\delta x'$ to the left given by eq. (23) before touching the proper top plane of the prism, assigning the second mark. A $xy$ instantaneous falling plane containing raindrops in $0xyz$ world becomes an inclined instantaneous falling plane in $0'x'y'z'$ world, being the inclination, eq. (21), easily derived from eqs. (23) and (25):
\begin{equation}
\tan{\left(\pi-\alpha\right)}=\frac{\delta z'(t')}{x'_{i+1}(t')-x'_{i}(t')}=\frac{\gamma u_{\infty}v}{c^{2}},
\end{equation}
giving the eq. (21).

The $t'$ instantaneous distance, $D_{c}$ along $0'x'$, between two consecutive inclined planes of raindrops crossing the proper top plane of the prism is given by\footnote{The vertical distance, along $0'z'$, between two consecutive planes of raindrops in $0'x'y'z'$ is $d_{0z}$, since two vertically located raindrops ($0z$ consecutivelly along), $t$ instantaneously in $0xyz$, have same $t$, same $x$ and $\delta z=d_{0z}$ . This implies the same $x'$, the same $t'$ and the same $\delta z'=d_{0z}$ Lorentz transformation for both in $0'x'y'z'$.}:
\begin{equation}
\tan{\left(\pi-\alpha\right)}=\frac{d_{0z}}{D_{c}}\stackrel{(26)}{\Rightarrow}D_{c}=\frac{d_{0z}c^{2}}{\gamma u_{\infty}v}.
\end{equation}
Hence, the $t'$ instantaneous fractional number of planes $n_{c}^{\pi}(t')$ crossing the proper top plane of the prism reads\footnote{See the drawing at page 1.}:
\begin{equation}
n_{c}^{\pi}(t')=\frac{AD}{D_{c}}\stackrel{(27)}{=}\frac{\gamma(AD)u_{\infty}v}{d_{0z}c^{2}}. 
\end{equation}
Raindrops belonging to one of these planes periodically cross, one by one, where:
\begin{equation}
T_{z'}=\frac{\delta z'}{\left(u_{\infty}/\gamma\right)}\stackrel{(23)}{=}\frac{\gamma v d_{0x}}{c^{2}}
\end{equation}
is the period. Hence, $n_{c}^{\pi}(t')$ is the $t'$ instantaneous number of raindrops periodically crossing the proper top surface of the prism with period $T_{z'}$ at an arbitrary instant $t'$ in a time interval $(t'_{0};\,\,t'_{0}+\delta t_{z'}^{\pi})$ such that:
\begin{equation}
\delta t_{z'}^{\pi}=\frac{d_{0z}}{\left(u_{\infty}/\gamma\right)}=\frac{\gamma d_{0z}}{u_{\infty}},
\end{equation}\\
\vspace{-0.5cm}
\\   
is the amount of time two vertically sucessive raindrops in two inclined raining planes crosses the proper top surface of the prism as observed in $0'x'y'z'$ world; $t'_{0}$ is the instant the first among these two sucessive raindrops touches the proper top surface of the prism as observed in $0'x'y'z'$ world.

Inserting the results of eqs. (4) and (5) in eq (3), one obtains:\\
\begin{displaymath}
\phi'_{r}=\int_{\textit{ABCDA}}\left(-v\hat{e}_{x'}-\frac{u_{\infty}}{\gamma}\hat{e}_{z'}\right)_{\textit{ABCDA}}\cdot\hat{n}dS'+\int_{\textit{ABGFA}}\left(-v\hat{e}_{x'}-\frac{u_{\infty}}{\gamma}\hat{e}_{z'}\right)_{\textit{ABGFA}}\cdot\hat{n}dS'\stackrel{(4),(5)}{\Rightarrow}\,\,\,\,\,\,\,\,\,\,\,\,\,\,\,\,\,\,\,\,\,\,\,\,
\end{displaymath}
\begin{displaymath}
\phi'_{r}=\int_{0}^{DA}\int_{0}^{AB}\left\{\left(-v\hat{e}_{x'}-\frac{u_{\infty}}{\gamma}\hat{e}_{z'}\right)\displaystyle\sum_{i=1}^{N_{x'}(t')}\sum_{j=1}^{N_{y'}(t')}\int_{-\infty}^{\infty}\int_{-\infty}^{\infty}\delta\left[x'-x_{i}'(t')\right]\delta\left[y'-y_{j}'(t)\right] \; dx'dy'\right\}\cdot\hat{e}_{z'} \; dx'dy' \; +
\end{displaymath}
\begin{displaymath}
\, \, \, \, \, \,  \,+\int_{0}^{FG}\int_{0}^{GB}\left\{\left(-v\hat{e}_{x'}-\frac{u_{\infty}}{\gamma}\hat{e}_{z'}\right)\displaystyle\sum_{j=1}^{N_{y'}(t')}\sum_{k=1}^{N_{z'}(t')}\int_{-\infty}^{\infty}\int_{-\infty}^{\infty}\delta\left[y'-y_{j}'(t')\right]\delta\left[z'-z_{k}'(t')\right] \; dy'dz'\right\}\cdot\hat{e}_{x'} \; dy'dz'=
\end{displaymath}
\begin{displaymath}
\,\,\,\,=\,\displaystyle\sum_{i=1}^{N_{x'}(t')}\sum_{j=1}^{N_{y'}(t')}-\frac{u_{\infty}}{\gamma}\int_{-\infty}^{\infty}\int_{-\infty}^{\infty}\left(\int_{0}^{DA}\int_{0}^{AB} \; dx'dy'\right)\delta\left[x'-x_{i}'(t')\right]\delta\left[y'-y_{j}'(t')\right] \; dx'dy' \; + \,\,\,\,\,\,\,\,\,\,\,\,\,\,\,\,\,\,\,\,\,\,\,\,\,\,\,\,
\end{displaymath}
\begin{displaymath}
\, \,\,\,\,\,-\,\displaystyle\sum_{j=1}^{N_{y'}(t')}\sum_{k=1}^{N_{z'}(t')}-v\int_{-\infty}^{\infty}\int_{-\infty}^{\infty}\left(\int_{0}^{FG}\int_{0}^{GB} \; dy'dz'\right)\delta\left[y'-y_{j}'(t')\right]\delta\left[z'-z_{k}'(t')\right] \; dy'dz'.\,\,\,\,\,\,\,\,\,\,\,\,\,\,\,\,\,\,\,\,\,\,\,\,\,\,\,\,\,\,\,\,\,\,\,\,\,\,\,\,\,\,\,\,
\end{displaymath}\\
One should infer that the double integrals between brackets, in the first and in the second integrals of the above expression, are the functions passing through the respective Dirac $\delta$ filters, returning, each, respectively, the projected areas of the raindrops passing through the respective surfaces of contact at the points $\left(x_{i}'(t'),y_{j}'(t')\right)$ and $\left(y_{j}'(t'),z_{k}'(t')\right)$. Denoting these projected areas by $a^{s}_{ij}$ e $a^{f}_{jk}$\footnote{Read footnote 2.}, one has:\\
\begin{equation}
\phi'_{r}=\displaystyle\sum_{i=1}^{N_{x'}(t')}\sum_{j=1}^{N_{y'}(t')}-\frac{u_{\infty}}{\gamma} a^{s}_{ij}+\displaystyle\sum_{j=1}^{N_{y'}(t')}\sum_{k=1}^{N_{z'}(t')}-va^{f}_{jk}.
\end{equation}\\
Writing down the contact flux $\phi'_{r}$, the instantaneous rate of variation of the raining water volume touching the prism while the prism runs through the rain to the finish line:\\
\begin{equation}
\frac{dV}{dt'}=\phi'_{r}=\displaystyle\sum_{i=1}^{N_{x'}(t')}\sum_{j=1}^{N_{y'}(t')}-\frac{u_{\infty}}{\gamma} a^{s}_{ij}+\displaystyle\sum_{j=1}^{N_{y'}(t')}\sum_{k=1}^{N_{z'}(t')}-va^{f}_{jk}.
\end{equation}\\
The variation of the proper volume $dV_{p}$ quantity entering the proper contact area of the prism in $0'x'y'z'$ world is related to the eq. (32) simply by:
\begin{displaymath}
\gamma^{-1}\frac{dV_{p}}{dt'}=\phi'_{r}=\displaystyle\sum_{i=1}^{N_{x'}(t')}\sum_{j=1}^{N_{y'}(t')}-\frac{u_{\infty}}{\gamma} a^{s}_{ij}+\displaystyle\sum_{j=1}^{N_{y'}(t')}\sum_{k=1}^{N_{z'}(t')}-va^{f}_{jk}\Rightarrow\,\,\,\,\,\,\,\,\,\,\,\,\,
\end{displaymath}
\begin{equation}
\frac{dV_{p}}{dt'}=\displaystyle\sum_{i=1}^{N_{x'}(t')}\sum_{j=1}^{N_{y'}(t')}-\frac{u_{\infty}}{\gamma} \left(\gamma a^{s}_{ij}\right)+\displaystyle\sum_{j=1}^{N_{y'}(t')}\sum_{k=1}^{N_{z'}(t')}-v\left(\gamma a^{f}_{jk}\right),
\end{equation}\\
\vspace{-0.4cm}
\\
and one easily verify the proper orthogonally projected areas, $a^{p,s}_{ij}$ and $a^{p,f}_{jk}$, of the raindrops through the proper top and frontal contact surfaces of the prism, rspectively, as being given by:
\begin{equation}
a^{p,s}_{ij}=\pi R_{0}^{2}=\gamma a^{s}_{ij}\Rightarrow a^{s}_{ij}=\gamma^{-1}\pi R_{0}^{2};
\end{equation} 
\begin{equation}
a^{p,f}_{jk}=\pi R_{0}^{2}=\gamma a^{f}_{jk}\Rightarrow a^{f}_{jk}=\gamma^{-1}\pi R_{0}^{2},
\end{equation}\\
\vspace{-0.55cm}
\\
where we denoted by $R_{0}$ the proper radius of the supposed spherical proper raindrops. Putting these last results in eq. (32), one has:
\begin{equation}
\frac{dV}{dt'}=-\pi R_{0}^{2}\left[\displaystyle\sum_{i=1}^{N_{x'}(t')}\sum_{j=1}^{N_{y'}(t')}\gamma^{-2}u_{\infty}+\displaystyle\sum_{j=1}^{N_{y'}(t')}\sum_{k=1}^{N_{z'}(t')}\gamma^{-1}v\right].
\end{equation}\\
\vspace{-0.25cm}
\\  
The minus sign is provided by our initial convention regarding the flux as being positive when exiting from a control surface. Hence the flux is negatively exiting the prism, i.e., entering through the prism. Summing up eq. (36):
\begin{equation}
\frac{dV}{dt'}=-\pi R_{0}^2\left[\gamma^{-2}u_{\infty} N_{x'}(t')N_{y'}(t')+\gamma^{-1}v N_{y'}(t')N_{z'}(t')\right],
\end{equation}\\
\vspace{-0.5cm}
\\
we must consider de eqs. (9) and (10).

Firstly, the period $T_{z'}$ is given by eq. (29). The period $T_{x'}$, the amount of time the prism spends to touch two consecutively spaced vertical planes of raindrops parallel to $y'z'$ in $0'x'y'z'$ world, reads:\\
\vspace{-0.5cm}
\\
\begin{equation}
T_{x'}=\frac{\gamma^{-1}d_{0x}}{v},
\end{equation}
since $\gamma^{-1}d_{0x}$ is the $t'$ instantaneous $0'x'$ along distance between two consecutive raindrops contained in a same inclined raindrop plane in $0'x'y'z'$, hence the distance beetween two consecutively spaced vertical (parallel to $y'z'$) planes in $0'x'y'z'$\footnote{See eq. (25). Also, one should attempt to the fact in the $0'x'y'z'$ world: to each raindrop contained in an inclined raindrop plane there exists another raindrop vertically above/below, being $d_{0z}$ de distance between them.}; $-v$ is the $0'x'$ (retrograd) velocity component of these planes in $0'x'y'z'$. The time duration blocks to be used in wich these periodic touchs hold is $\delta t_{z'}^{\pi}$, given by eq. (30). The summation limits in eqs. (9) and (10) are given by:
\begin{equation}
I\left(\tau'/T_{z'}\right)\stackrel{(29),(30)}{=}I\left(\frac{c^{2}}{u_{\infty}v}\frac{d_{0z}}{d_{0x}}\right);
\end{equation}
\begin{equation}
I\left(\tau'/T_{x'}\right)\stackrel{(30),(38)}{=}I\left(\frac{\gamma^{2}v}{u_{\infty}}\frac{d_{0z}}{d_{0x}}\right).\,\,\,
\end{equation}\\
\vspace{-0.5cm}
\\
In virtue of eq. (37), we easily obtain from the eqs. (9) and (10):
\begin{equation}
\gamma^{-2}\pi R_{0}^{2}u_{\infty}N_{x'}(t')N_{y'}(t')=\frac{4}{3}\pi\gamma^{-1}R_{0}^{3}\sum_{k=0}^{I\left(\tau'/T_{z'}\right)}\,\,\,\,N_{x'}N_{y'}\delta\left(t'-k\frac{\gamma vd_{0x}}{c^{2}}\right);\,\,\,\,
\end{equation}
\begin{equation}
\gamma^{-1}\pi R_{0}^{2}v N_{y'}(t')N_{z'}(t')=\frac{4}{3}\pi\gamma^{-1}R_{0}^{3}\sum_{l=0}^{I\left(\tau'/T_{x'}\right)}N_{y'}N_{z'}\delta\left(t'-l\gamma^{-1}\frac{d_{0x}}{v}\right).
\end{equation}
The spent time, in $0'x'y'z'$, to cross the finish line is given by the instant in wich the origin $0'$ crosses the proper distance $L$ measured in $0xyz$:
\begin{equation} 
t'_{L}-0=\gamma\left(\frac{L}{v}-\frac{v}{c^{2}}L\right)=\gamma \frac{L}{v}\left(1-\frac{v^{2}}{c^{2}}\right)=\gamma^{-1}\frac{L}{v},
\end{equation}
being the Lorentz contraction of the spent time, $L/v$, in $0xyz$.
From eqs. (37), (41), (42) and (43), the total non-proper volume $V_{L}$, passing through a transparently adjacent proper contact surfaces of the prism is given by:
\begin{equation}
V_{L}=\frac{4}{3}\pi\gamma^{-1}R_{0}^{3}\int_{0}^{\gamma^{-1}L/v}\left[\sum_{k=0}^{I\left(\tau'_{L}/T_{z'}\right)}N_{x'}N_{y'}\delta\left(t'-k\frac{\gamma vd_{0x}}{c^{2}}\right)+\sum_{l=0}^{I\left(\tau'_{L}/T_{x'}\right)}N_{y'}N_{z'}\delta\left(t'-l\gamma^{-1}\frac{d_{0x}}{v}\right)\right]dt',
\end{equation}
where the time duration blocks are now summed up to $\tau'_{L}=\gamma^{-1}L/v$. Since the Dirac delta filters raining raindrops in the time domain related to the proper lenght $L$, one extends de integration domain with the condition: $N_{x'}N_{y'}$ and $N_{y'}N_{z'}$ are void in the interval $t'$ $\notin$ $[0;\,\gamma^{-1}L/v]$. Hence, the eq. (44) reads\footnote{When one extends the domain, it should be asseverated that the Dirac delta is under a distribution context. The raindrops filter process is strongly concentrated, but not instantaneous. Hence, the totality of raindrops being filtered by the integration over the Dirac delta summation must be properly treated in the left hand side of eq. (44) regarding the volume integration. If one states that none raindrop is being filtered by the delta at $t'=0$ and that at $t=\gamma^{-1}L/v$ all raindrops are properly filtered, one shall perform the delta summation starting from $k=1$ and, $V_{L}$, \textit{the volume variation}, straighforwardly gives the totality of filtered raindrops. If raindrops start to be filtered at $t'=0$, as we wrote down, the volume integration must be rigorously understood as (e.g., the first filtering process) $V_{L}|_{t'=\gamma^{-1}L/v}-V_{L}|_{t'=0}=\lim_{\epsilon\rightarrow 0}\left(\int_{0}^{0+\epsilon}dV+\int_{\epsilon}^{\gamma^{-1}L/v}dV\right)=\left(4/3\right)\pi\gamma^{-1}R_{0}^{3}+\int_{0}^{\gamma^{-1}L/v}dV=\left(4/3\right)\pi\gamma^{-1}R_{0}^{3}+V_{L}$. This process leads one to conclude the $k$ summation indice must starts from $1$ in any case.}:
\begin{equation}
V_{L}=\frac{4}{3}\pi\gamma^{-1}R_{0}^{3}\left[\sum_{k=1}^{I\left(\tau'_{L}/T_{z'}\right)}\int_{-\infty}^{\infty}N_{x'}N_{y'}\delta\left(t'-k\frac{\gamma vd_{0x}}{c^{2}}\right)dt'+\sum_{l=1}^{I\left(\tau'_{L}/T_{x'}\right)}\int_{-\infty}^{\infty}N_{y'}N_{z'}\delta\left(t'-l\gamma^{-1}\frac{d_{0x}}{v}\right)dt'\right].
\end{equation}
In virtue of eq. (11), one writes:
\begin{displaymath}
V_{L}=\frac{4}{3}\pi\gamma^{-1} R_{0}^{3}\left[\sum_{k=1}^{I\left(\tau'_{L}/T_{z'}\right)}N_{s}\left(k\frac{\gamma vd_{0x}}{c^{2}}\right)+\sum_{l=1}^{I\left(\tau'_{L}/T_{x'}\right)}N_{f}\left(l\gamma^{-1}\frac{d_{0x}}{v}\right)\right]\Rightarrow
\end{displaymath}
\begin{equation}
V_{L}=\frac{4}{3}\pi\gamma^{-1} R_{0}^{3}\left[I\left(\gamma^{-2}\frac{c^{2}}{v^{2}}\frac{L}{d_{0x}}\right)N_{s}+I\left(\frac{L}{d_{0x}}\right)N_{f}\right].\,\,\,\,\,\,\,\,\,\,\,\,\,
\end{equation}\\
\vspace{-0.4cm}
\\
From eq. (8) and (27)\footnote{Remembering: $AT=d_{0z}$ and $AS=D_{c}$.}, the eq. (46) becomes:
\begin{equation}
V_{L}=\frac{4}{3}\pi\gamma^{-1}R_{0}^{3}u_{\infty}N_{f}\left[\frac{1}{v}I\left(\gamma^{-2}\frac{c^{2}}{v^{2}}\frac{L}{d_{0x}}\right)\gamma\frac{v^{2}}{c^{2}}\frac{A_{s}}{A_{f}}+\frac{1}{u_{\infty}}I\left(\frac{L}{d_{0x}}\right)\right].
\end{equation}
Hence, the proper volume accumulating upon the proper prism surface is given by:
\begin{equation}
V_{L}^{p}=\frac{4}{3}\pi R_{0}^{3}u_{\infty}N_{f}\left[\frac{1}{v}I\left(\gamma^{-2}\frac{c^{2}}{v^{2}}\frac{L}{d_{0x}}\right)\gamma\frac{v^{2}}{c^{2}}\frac{A_{s}}{A_{f}}+\frac{1}{u_{\infty}}I\left(\frac{L}{d_{0x}}\right)\right],
\end{equation}
or the number of raindrops, $V_{L}^{p}/(4\pi R_{0}^{3}/3)$:
\newpage
\begin{equation}
N_{L}=N_{f}\left[\frac{u_{\infty}}{v}I\left(\gamma^{-2}\frac{c^{2}}{v^{2}}\frac{L}{d_{0x}}\right)\gamma\frac{v^{2}}{c^{2}}\frac{A_{s}}{A_{f}}+I\left(\frac{L}{d_{0x}}\right)\right].
\end{equation}
Let's analyze the classical limit. Taking the limit $c\rightarrow\infty$, one obtains the classical (galilean) non-relativistic number of raindrops $N_{L}^{c}$:
\begin{equation}
N_{L}^{c}=N_{f}I\left(\frac{L}{d_{0x}}\right)\left(\frac{u_{\infty}}{v}\frac{A_{s}}{A_{f}}+1\right),
\end{equation}
being the assintotically minimum contact, $N_{f}L/d_{0x}$, at $v\rightarrow\infty$. Relativistically (Einstein's), the same minimum amount of contact is reached at $v=c$, but not assintotically. One stays more dry by running faster, but in a Lorentzian world one always stays less wet.
\end{document}